\documentclass[11pt, a4paper]{article}
\pdfoutput=1
\usepackage{jcappub}

\usepackage{comment}
\usepackage{amsmath}
\usepackage{amsfonts}
\usepackage{amssymb}
\usepackage{graphicx}
\usepackage{mathrsfs}
\usepackage{latexsym}
\usepackage{color}
\usepackage{slashed, cancel}
\usepackage{hyperref}
\usepackage{cleveref}
\usepackage{float}
\usepackage{tikz}
\usepackage{bbold}
\usepackage{soul} 
\usepackage{mathtools}
\usepackage{support-caption} 
\usepackage{subcaption}
\usepackage{enumitem}
\usepackage{journals}
\allowdisplaybreaks
\usepackage[version=3]{mhchem}
\hypersetup{urlcolor=blue, colorlinks=true} 

\usepackage{fancyhdr}
\numberwithin{equation}{section}

\definecolor{orange}{rgb}{1,0.4,0}
\definecolor{green}{rgb}{0,0.65,0}
\definecolor{rossos}{rgb}{0.8,0.2,0.3}
\definecolor{bluscuro}{rgb}{0.15, 0.2, .85}
\definecolor{bluchiaro}{cmyk}{1,.3,0.,0.1}
\hypersetup{colorlinks, citecolor=bluscuro, linkcolor=bluscuro, urlcolor=bluscuro}

\makeatother   

\begin{document}

\today

 \title{$Z^\prime$-mediated dark matter 
 freeze-in  \\ ~~~~~~~~~~at stronger coupling}

\author[a,b]{Giorgio Arcadi,}
\author[a,b,c,d]{David Cabo-Almeida,}
\author[e]{Oleg Lebedev }

\affiliation[a]{Dipartimento di Scienze Matematiche e Informatiche, Scienze Fisiche e Scienze della Terra, \\ Universita degli Studi di Messina, Viale Ferdinando Stagno d'Alcontres 31, I-98166 Messina, Italy}
\affiliation[b]{INFN Sezione di Catania, Via Santa Sofia 64, I-95123 Catania, Italy}
\affiliation[c]{Departament de Física Quàntica i Astrofísica, Universitat de Barcelona,
Martí i Franquès 1,\\ E08028 Barcelona, Spain}
\affiliation[d]{Institut de Ciències del Cosmos (ICCUB), Universitat de Barcelona,
Martí i Franquès 1,\\ E08028 Barcelona, Spain}
\affiliation[e]{Department of Physics and Helsinki Institute of Physics,
Gustaf H\"allstr\"omin katu 2a, \\
FI-00014 Helsinki, Finland}

\emailAdd{giorgio.arcadi@unime.it}
\emailAdd{david.cabo@ct.infn.it}
\emailAdd{oleg.lebedev@helsinki.fi}

\abstract{We study freeze-in production of fermionic dark matter mediated by a $Z^\prime$ gauge boson. In particular, we explore the regime of Boltzmann-suppressed production, when the Standard Model (SM) thermal bath
temperature never exceeds the dark matter mass. The corresponding  gauge coupling is then required to be significant, up to order one. As a result, this class of freeze-in models can be probed by the current and future direct  
dark matter detection experiments. }

\maketitle


  \section{Introduction}
   
   Cosmological dark matter remains one of the most mysterious entities in modern physics. The most common approach to this problem is to postulate the existence of a stable particle which has no quantum numbers with respect to the Standard Model
   gauge symmetries and whose gravitational effects we observe. Its further properties depend on the coupling to the observed particles as well as itself. If such couplings are very small, the dark matter field never reaches thermal equilibrium and its abundance accumulates over time due to 
   its slow production by the SM thermal bath \cite{Dodelson:1993je}. This  mechanism is known as ``freeze-in'' \cite{Hall:2009bx}.
   
   The freeze-in mechanism is predictive only if other production channels are negligible. Generally, there are further non-thermal DM production mechanisms which are operative in the Early Universe \cite{Lebedev:2022cic}. Most importantly, all particles are produced by gravity itself during and after inflation. For example, particle production immediately after inflation, during the inflaton oscillation epoch, can be very efficient due to the presence of the Planck-suppressed operators that couple the inflaton  $\phi$ to DM such as \cite{Lebedev:2022ljz,Koutroulis:2023fgp}
   \begin{equation}
{1\over M_{\rm Pl}^2 } \, \phi^4 s^2 ~,~ {1\over M_{\rm Pl} } \, \phi^2 \bar \chi\chi \;,
\end{equation}
where $s$ is scalar DM and $\chi$ is fermion DM.
   These operators are generated by both classical and quantum gravitational effects. The resulting abundance of DM exceeds the observed value by many orders of magnitude unless the corresponding Wilson coefficients happen to be very small or DM is extremely light
    \cite{Lebedev:2022cic}.
   
   One way to address this problem is to allow for a long inflaton-dominated expansion period before reheating. The  DM quanta produced immediately after inflation  are relativistic and their energy density gets diluted by expansion in a non-relativistic background. At the reheating stage, the 
   resulting DM abundance  is then small  \cite{Lebedev:2022cic}.
   This implies  that the
   reheating temperature $T_R$ is  relatively low, being only limited from below, $T_R > 4\;$MeV,  by observations \cite{Hannestad:2004px}. It opens the possibility that the SM bath temperature has never been higher than the DM mass and the freeze-in production is Boltzmann suppressed \cite{Cosme:2023xpa}.
   The SM-DM coupling required for producing the correct DM relic abundance can then be as large as order one, which facilitates direct searches for dark matter   via DM-nucleon scattering and  in collider experiments.

   The purpose of our current work is to explore this idea in the context of a $Z^\prime$ extension of the Standard Model \cite{Langacker:2008yv}. The role of dark matter is played by a Dirac fermion which couples to a $Z^\prime$ and whose production by the SM thermal bath is mediated by the
   $Z^\prime$. We delineate parameter space leading to the correct DM relic abundance, while satisfying   the current direct detection and collider constraints. We find that the $Z^\prime$-mediated freeze-in can be probed by direct DM detection experiments as long as
   the $Z^\prime$ has significant ``vector'' couplings to the SM fermions.

   \subsection{The model}
   
   We study an extension of the Standard Model with a massive $Z^\prime$ boson and a Dirac fermion $\chi$, which has no SM quantum numbers and plays the role of dark matter. The relevant couplings are
   \begin{equation}
\label{eq:lagmicro}
\mathcal{L} \supset-m_\chi \bar{\chi} \chi-\frac{1}{2} M^2_{Z^{\prime}} Z_\mu^{\prime} Z^{\prime \mu}+\bar{\chi} \gamma^\mu\left(V_\chi-A_\chi \gamma_5\right) \chi \;Z_\mu^{\prime}+\sum_f \bar{f} \gamma^\mu\left(V_f-A_f \gamma_5\right) f \;Z_\mu^{\prime}\;,
\end{equation}
Here $f$ represents the SM fermions;  $V_{f,\chi}$ and $A_{f,\chi}$ are the vectorial and axial couplings. These are related to the $Z^\prime$ gauge coupling  $g_{Z^\prime}$  and the charges of the left- and right-handed fermions   $X_{f_L},X_{f_R}$      as 
 $V_f=g_{Z^\prime}(X_{f_L}+X_{f_R})/2$ and $A_f=g_{Z^\prime}(X_{f_L}-X_{f_R})/2$.

   We assume that the $Z^\prime $ is the only portal between the SM and dark matter, focussing on the TeV scale $M_{Z^{\prime}}$. 
      We also  take its mixing with the SM gauge bosons to be negligible for our purposes. The U(1) extensions of the SM are subject to the anomaly cancellation constraint, which generally necessitates the presence of additional fermions. Such fermions, however, can be vector-like with respect to the SM gauge charges and hence  have large masses making their impact on the DM phenomenology insignificant. A similar approach was taken in related work \cite{Arcadi:2013qia,Lebedev:2014bba} (see also \cite{Arcadi:2014lta}).

   In what follows, we study freeze-in production of dark matter $\chi$ mediated by the $Z^\prime $. Our main assumption is that the SM thermal bath temperature $T$ is always below $m_\chi$. This means, in particular, that the reheating temperature $T_R$ is sufficiently low,
   allowing for dilution of the gravitationally produced relics, as explained above.
   To compute the DM production, we resort to the instant reheating approximation, e.g. take  $T$ to increase abruptly from zero to $T_R$ and then decrease as usual, in accordance with the SM entropy conservation. 
   This gives a good estimate of the DM abundance in cosmological models with a flat SM temperature profile before reheating \cite{Cosme:2024ndc}. 
   Using appropriate rescaling and replacing $T_R$ with the maximal SM sector temperature, we can extend our results to a broader  class of models.             
   Earlier work on the $Z^\prime$-mediated freeze-in, which assumes a high reheating temperature, can be found in \cite{Cosme:2021baj}.

\section{Dark matter  relic abundance}
   
   In this section, we study freeze-in dark matter production in the Boltzmann-suppressed regime $T\ll m_\chi$. 
   Dark matter  is produced via annihilation of the SM thermal bath particles, e.g.  $\bar f f \rightarrow Z^\prime \rightarrow \bar \chi \chi$, and its density $n= n_\chi + n_{\bar \chi}$ is controlled by the Boltzmann equation,
   \begin{equation}
   \dot n + 3Hn =2\, \Gamma ( {\rm SM}\rightarrow \bar \chi \chi) - 2\, \Gamma ( \bar \chi \chi \rightarrow  {\rm SM} ) \;,
   \label{Boltzmann}
   \end{equation}
   where the factor of 2 accounts for production of 2 DM states, $\chi$ and $\bar \chi$, and $\Gamma$ is the reaction rate per unit volume.
   
    Since dark matter is non-relativistic at $T \ll m_\chi$, the $a\rightarrow b$ reaction rate is given by 
\begin{equation}
\Gamma_{a\rightarrow b} = \int \left( \prod_{i\in a} {d^3 {\bf p}_i \over (2 \pi)^3 2E_{i}} f(p_i)\right)~
\left( \prod_{j\in b} {d^3 {\bf p}_j \over (2 \pi)^3 2E_{j}} \right)
\vert {\cal M}_{a\rightarrow b} \vert^2 ~ (2\pi)^4 \delta^4(p_a-p_b) \;,
\label{Gamma}
\end{equation}
where $p_i$ and $p_j$ are the initial and final state momenta, respectively, and $f(p)$ is the momentum distribution function. 
${\cal M}_{a\rightarrow b}$ is the  QFT  $a \rightarrow b$  transition amplitude, 
in which 
  both
 the { initial and final} state phase space symmetry factors are absorbed.
For the $2\rightarrow 2$ reactions,  energy conservation combined with the Boltzmann statistics $f(p)=e^{-E/T}$ implies
\begin{equation}
f (p_1) f(p_2) = f(p_3) f (p_4)\;.
\end{equation}
   Therefore, the DM production rate via thermal SM states can be written as the  {\it thermal} DM annihilation rate into SM quanta,
   \begin{equation}
 \Gamma ( {\rm SM}\rightarrow \bar \chi \chi) = \Gamma ( \bar \chi \chi \big\vert_{\rm therm}  \rightarrow {\rm SM}) \;.
 \label{eq0}
\end{equation}
 The latter can be computed following Gelmini and Gondolo
   \cite{Gondolo:1990dk},
 \begin{eqnarray}
  \Gamma (\bar \chi \chi \big\vert_{\rm therm}  \rightarrow {\rm SM })&=& \langle \sigma(\bar \chi \chi \big\vert_{\rm therm}  \rightarrow {\rm SM } )\,   v_r \rangle  \, n_\chi n_{\bar \chi} = {2^2\over (2\pi)^6} \;     \int \sigma v_r \, e^{-E_1/T} e^{-E_2/T}  d^3 p_1 d^3 p_2  \nonumber \\
  &=&  {2^3\pi^2 T\over (2\pi)^6}\;
 \int_{4m_\chi^2}^\infty  d{\rm s} \; \sigma \;({\rm s}- 4 m_\chi^2) \sqrt{{\rm s}} \, K_1 (\sqrt{{\rm s}}/T) \;,
 \label{Gammahhss}
 \end{eqnarray}
 where $\sigma$ is the   $\bar \chi \chi    \rightarrow {\rm SM } $    cross section; $v_r$ is the relative velocity of the colliding quanta with energies $E_1,E_2$ and momenta $p_1,p_2$; $n_\chi$ is the $\chi$  number density;
 $\langle ... \rangle $ denotes a thermal average;
 ${\rm s}$ is the Mandelstam variable, and
 $K_1(x)$ is the modified Bessel function of the first kind.   We have explicitly factored out the spin d.o.f. factor $2^2$ specific to fermion annihilation  ($n_\chi= 2\, e^{-E/T}$).

   Let us  focus for now on a heavy $Z^\prime$ regime, $M_{Z' } > 2m_\chi$. It is instructive to consider in detail the limit $M_{Z' } \gg m_\chi, m_f$, where $f$ represents the SM fermions. This allows us to obtain simple analytical approximations in the pure freeze-in limit. In our numerical studies however, we do not resort to this approximation and use the exact (tree-level) results together with the backreaction term in (\ref{Boltzmann}).
In what follows, we  consider the vector and axial $Z^\prime$ cases separately.

  \subsection{Vector coupling}
  
  In most of the parameter space, the main production/annihilation channel is 
  \begin{equation}
  \bar \chi \chi      \leftrightarrow \bar f f \,,
  \end{equation}
  where $f$ is an SM fermion and contributions of all the fermions lighter than $\chi$ should be summed.
  The DM annihilation via a vector $Z^\prime$ is allowed already at the $s$-wave level and the corresponding cross section for $m_\chi \gg m_f$ is 
  \begin{equation}
  \sigma ({\bar \chi} \chi \rightarrow \bar f f) \simeq  {V_f^2 V_\chi^2  m_\chi^2 \over 2\pi M_{Z'}^4} \, \sqrt{s\over  s-4m_\chi^2}  \;.
  \end{equation}
  The reaction rate is found from Eq.\,\ref{Gammahhss}.
  In the regime $m_\chi \gg T$, we may use the asymptotic form $K_1 (\sqrt{{\rm s}}/T) \simeq \sqrt{\pi\over 2} {T^{1/2}\over s^{1/4} } e^{-\sqrt{s} /T }$. 
   When computing the integral over $s$, one can use the following approximation: since the integrand peaks sharply close to $s= 4m_\chi^2$, one may replace
  $s\rightarrow  4m_\chi^2$ in ``slow'' functions of $s$, while keeping the factors $\sqrt{  s-4m_\chi^2}$ and $e^{-\sqrt{s} /T}$ as they are. 
   The resulting integral reduces to the Gamma function such that
   \begin{equation}
  \int_{4m_\chi^2}^\infty  d{\rm s} \;  \sqrt{  s-4m_\chi^2} \; e^{-\sqrt{s} /T} \simeq {\sqrt{\pi} \over 2} (4 m_\chi T)^{3/2} e^{-2m_\chi /T}\;.
  \label{int-Gamma}
  \end{equation}
   The resulting DM production rate via SM fermion annihilation  according to (\ref{eq0}) is
   \begin{equation}
 \Gamma (\bar f f \rightarrow \bar \chi \chi) = {V_f^2 V_\chi^2 \;  m_\chi^5 T^3 \over 2\pi^4 M_{Z'}^4} \, e^{-2m_\chi /T}\;.
  \end{equation}
   
   In the pure freeze-in regime,  the Boltzmann equation (\ref{Boltzmann}) reduces to $\dot n + 3nH =2 \,  \Gamma (\bar f f \rightarrow \bar \chi \chi) $. It can be solved analytically for simple reheating scenarios. For definiteness, 
   we resort to the instant reheating approximation, that is, we assume that the SM sector temperature increases abruptly from zero to $T_R$. After that it is determined by entropy conservation, as usual.
   Integrating the Boltzmann equation from $T=T_R$ to $T=0$, one finds that $n\propto T^3$ at late times and, due to the exponential suppression of the reaction rate,  the production is dominated by the initial moments when $T \simeq T_R$.
   Using
   \begin{equation}
s_{\rm SM}= {2\pi^2 \over 45} g_* T^3~,~ H = \sqrt{g_* \pi^2 \over 90 } \, {T^2 \over M_{\rm Pl}} \;,
\end{equation}
  where $g_*$ is the effective number of the  SM d.o.f. and $M_{\rm Pl}= 2.4 \times 10^{18}{\rm GeV}$, 
   we find
   \begin{equation}
   Y \equiv {n\over s_{\rm SM}} \simeq \sum_f {45 \sqrt{90} \over 4}  {V_f^2 V_\chi^2 \over \pi^7 g_*^{3/2}} \,{m_\chi^4 M_{\rm Pl} \over  M_{Z'}^4 T_R}  \, e^{-2m_\chi /T_R}\;,
   \end{equation}
   where the sum runs over all SM Dirac fermions $f$  with the color multiplicity properly included.\footnote{The neutrinos contribute 1/2 of the Dirac fermion contribution. For $V_f = V_\chi =\lambda$ and $m_\chi > m_t$,
   the sum over all the SM fermions amounts to a factor of 22.5.}
   Here we have assumed that   $g_*$ stays approximately constant in the regime of interest.
   For the universal couplings $V_f = V_\chi =\lambda$, the observed $Y=4.4 \times 10^{-10}\, {\rm GeV}/m_\chi$ requires
   \begin{equation}
    \lambda \simeq  10^{-6} \; {M_{Z'} T_R^{1/4} \over m_\chi^{5/4} } \, e^{m_\chi/(2T_R)} 
    \label{lambda}
       \end{equation}
   for $g_* \simeq 107$.
   Since $m_\chi \gg T_R$, the exponential factor can be  large making $\lambda$ as large as ${\cal O} (1) $.  Nevertheless, the produced dark matter density is low and it does not thermalize, thereby justifying the assumption of 
   the  freeze-in regime.  We note that the size of the  coupling is determined primarily by   $m_\chi /T_R$ and rather insensitive to a specific charge assignment, thus making the  universal charge limit justified.

   The above  result is obtained in the instant reheating approximation. As shown in \cite{Cosme:2024ndc},  this gives a good estimate of the required coupling within  a larger class of models. If the SM temperature is constant before reheating, 
     Eq.\,\ref{lambda} stands as it is, up to a small (percent level) shift in $T_R$. More generally, DM production peaks at the maximal temperature $T_{\rm max}$ such that one replaces $T_R \rightarrow T_{\rm max}$ in the above formulas
     and, to account for entropy production between $T=T_{\rm max}$ and $T=T_R$, rescales the coupling with the factor $(T_R/T_{\rm max})^\kappa$, where $\kappa$ is model dependent. 
   
   The common feature of freeze-in models at stronger coupling is that the DM abundance exhibits the factor $e^{-2m_{\rm DM} /T_R}$ times some power of the dark matter coupling \cite{Koivunen:2024vhr,Arcadi:2024wwg}. 
   As in the other models, we find that an order one coupling requires $m_{\rm DM} \sim 20 \,T_R$ for typical parameter values.

    \subsection{Axial coupling}
   
   Consider now DM production via  $\bar f f \rightarrow \bar \chi \chi$ mediated by an axial $Z^\prime$ \cite{Lebedev:2014bba}.
   To compute the reaction rate, we consider non-relativistic DM annihilation. In contrast to the previous case, in the $m_f \rightarrow 0$ limit, the process is $p$-wave and hence less efficient.
   Neglecting the fermion masses, one finds
    \begin{equation}
  \sigma ({\bar \chi} \chi \rightarrow \bar f f) \simeq  {A_f^2 A_\chi^2  \over 12 \pi M_{Z'}^4} \, \sqrt{s(  s-4m_\chi^2)}  \;.
  \end{equation}
   This approximation applies for heavy DM, $m_\chi \gg m_t$, or  in the regime where the top-quark channel is not available, $m_\chi < m_t$.
   The reaction rate calculation proceeds as before, except the different velocity dependence leads to the integral 
    \begin{equation}
  \int_{4m_\chi^2}^\infty  d{\rm s} \;  (s-4m_\chi^2)^{3/2} \; e^{-\sqrt{s} /T} \simeq  {3\sqrt{\pi} \over 4} (4 m_\chi T)^{5/2} e^{-2m_\chi /T}
  \end{equation}
   instead of (\ref{int-Gamma}).
   The result is
    \begin{equation}
 \Gamma (\bar f f \rightarrow \bar \chi \chi) = {A_f^2 A_\chi^2 \;  m_\chi^4 T^4 \over 2\pi^4 M_{Z'}^4} \, e^{-2m_\chi /T}\;.
  \end{equation}
   Note that the rate is now proportional to $T^4$, unlike that in the vector case.
   The DM abundance is given by
   \begin{equation}
   Y \simeq \sum_f {45 \sqrt{90} \over 4} {A_f^2 A_\chi^2 \over \pi^7 g_*^{3/2}} \,{m_\chi^3 M_{\rm Pl} \over  M_{Z'}^4 }\, e^{-2m_\chi /T_R} \;.
   \end{equation}
For $A_f = A_\chi =\lambda$ and $g_* \simeq 107$, the required coupling is  
\begin{equation}
    \lambda \simeq   10^{-6} \; {M_{Z'}  \over m_\chi } \, e^{m_\chi/(2T_R)} \;.
       \end{equation}

             \subsection{Additional channels}
        
      For heavier DM, $m_\chi \sim M_{Z'}$, the decay and $t$-channel processes 
      \begin{equation}
      Z^\prime\rightarrow \bar \chi \chi ~,~ Z^\prime Z^\prime \rightarrow \bar \chi \chi
      \end{equation}
        become important.
  The $Z^\prime $ number density receives  Boltzmann suppression similar  to $e^{-2m_\chi /T}$, hence processes with the $Z^\prime $ on-shell cannot be neglected.
      The $Z^\prime$ gauge boson  maintains  thermal equilibrium  with the SM bath due to its coupling to light fermions. The reaction
      \begin{equation}
      \bar f f \leftrightarrow Z^\prime
      \end{equation}
      is very efficient for the coupling range of interest and faster than the rate of expansion of the Universe (see e.g. analogous calculations in \cite{Arcadi:2024wwg}), thereby implying thermal equilibrium. 
      
\begin{figure}[h!]
    \centering
    \includegraphics[width=0.48\textwidth]{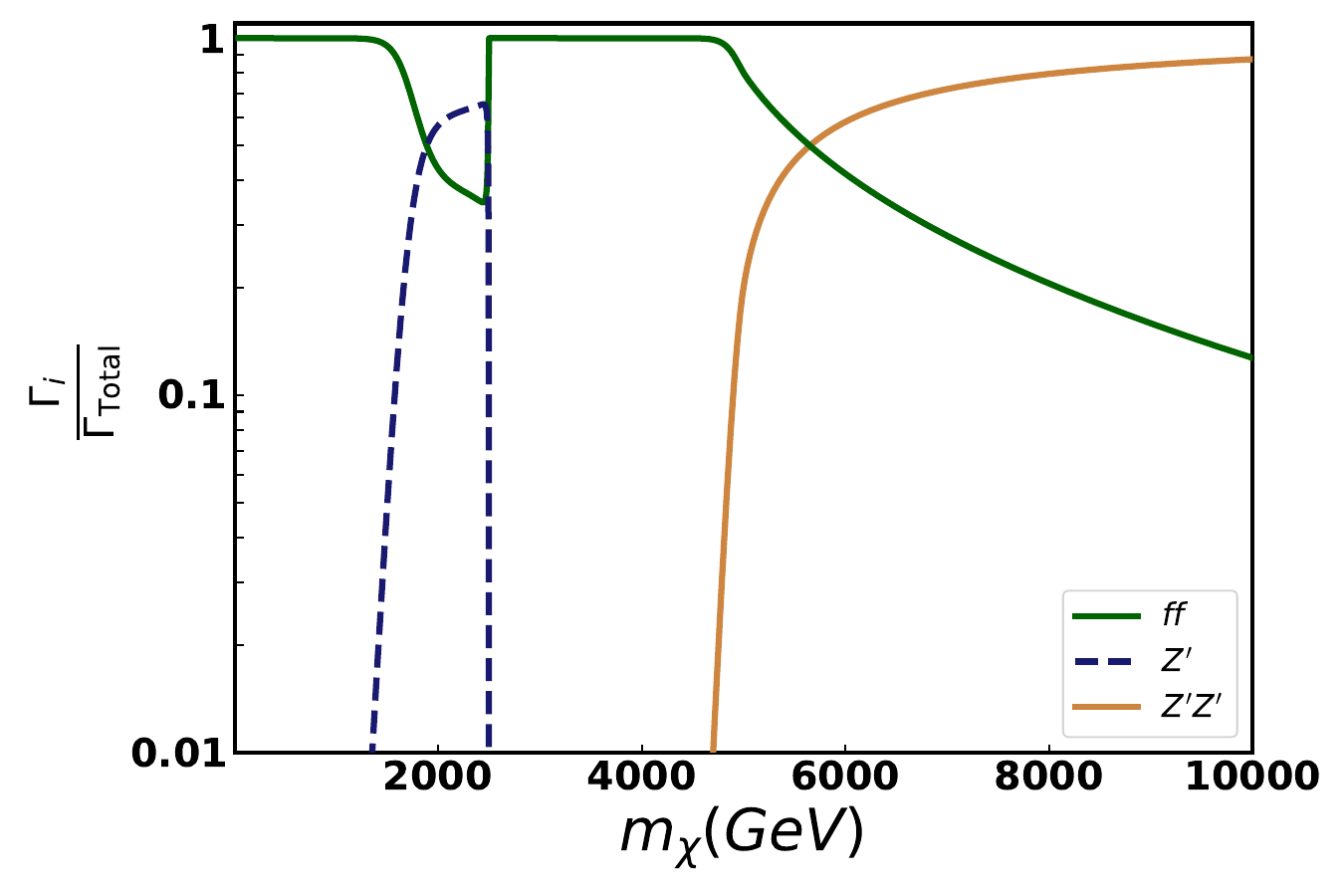}
    \includegraphics[width=0.5\textwidth]{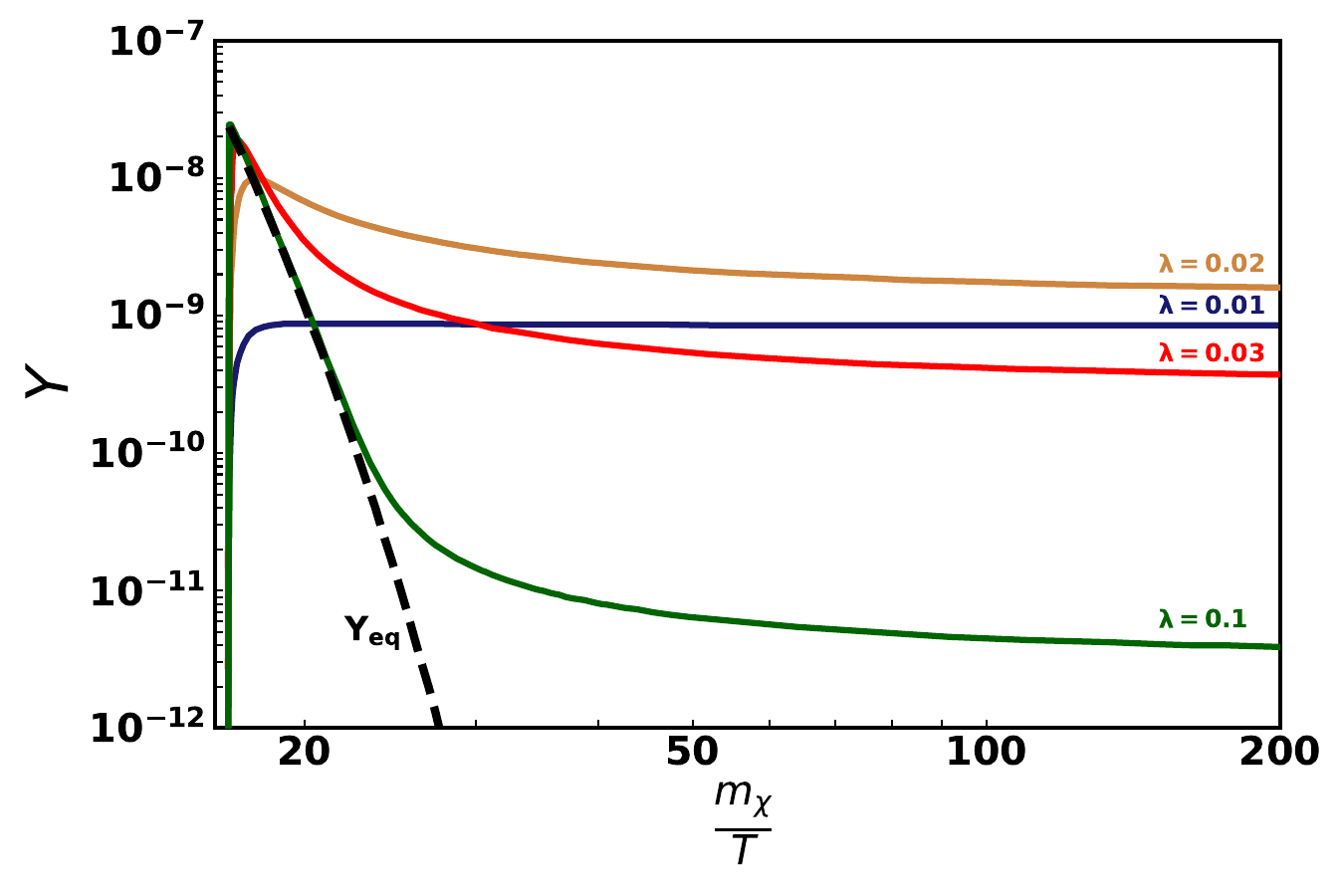}
    \caption{ {\it Left:} Dark matter production  relative  contributions $\Gamma_i / \Gamma_{\rm tot}$ for the $s$-channel, $t$-channel and decay reactions. Here the $axial$ coupling is assumed: $ A_f =A_\chi=0.1$ ($V_f=V_\chi=0$);  $M_{Z'}= 5$\, TeV and $T_R=120\;$GeV. ~
    {\it Right:} Evolution of the DM abundance $Y$ for different couplings. 
    Here $m_\chi =300\;$GeV, $M_{Z'}=900\; $GeV and $T_R=18\;$GeV; the coupling is defined by 
    $\lambda= V_f=V_\chi = A_f =A_\chi$. $Y_{\rm eq}$ represents the thermal equilibrium value.}
    \label{lambda-example}
\end{figure}

      The decay channel $Z^\prime\rightarrow \bar \chi \chi$ dominates  around $M_{Z'} \sim 2m_\chi$, while the efficiency of the $t$-channel mode $Z^\prime Z^\prime \rightarrow \bar \chi \chi$ depends on the nature of
      the $Z^\prime$ couplings. In case of the axial couplings, the fermion annihilation mode is suppressed by $m_f$ or DM velocity. On the other hand, the processes involving the longitudinal component of a $Z^\prime $
      at high energy are enhanced by $E/M_{Z'}$. Hence, one expects  $Z^\prime Z^\prime \rightarrow \bar \chi \chi$ to dominate for large $m_\chi$. 
      Fig.\,\ref{lambda-example} (left) shows  the relative reaction rate contributions of the different production channels for the axial coupling case, produced with   {\tt micrOMEGAs}  \cite{Belanger:2018ccd,Alguero:2023zol}.
      As expected, we observe that the $t$-channel annihilation becomes dominant for heavy dark matter.  In the vector coupling case, on the other hand, 
       the SM fermion annihilation mode is efficient and we find that it remains the main  production channel at large $m_\chi$.

   \subsection{Dark matter thermalization}

    The above  considerations help us understand the qualitative behaviour of  $\lambda (m_\chi, T_R)$  at low DM densities, when the term $\Gamma ( \bar \chi \chi \rightarrow  {\rm SM} ) $ in Eq.\,\ref{Boltzmann} can be neglected.
    However, in our numerical analysis we use the {\tt micrOMEGAs} tool  \cite{Belanger:2018ccd,Alguero:2023zol}
      which takes into account all the channels 
      as well as the DM annihilation effects. 
      The latter lead to a qualitative change in the relic abundance calculations at larger coupling.
        As one increases $\lambda$, the DM density grows thus enhancing DM annihilation and eventually leading to its thermalization.
          This manifests itself in the freeze-in lines  $\lambda (m_\chi, T_R)$   merging with the freeze-out curve
            $\lambda (m_\chi)$, as we show in the next section. Such thermalization   has been studied in detail in \cite{Cosme:2023xpa} and entails a smooth FIMP-WIMP transition \cite{Silva-Malpartida:2024emu}.

Fig.\,\ref{lambda-example} (right)  illustrates the effect of the coupling increase on the eventual DM abundance. While at small $\lambda$, $Y$ grows with the coupling,  this ceases to be the case above a certain critical coupling, which is around $\lambda \sim 0.02$ in this example.
We observe that the abundance decreases over time due to DM annihilation.
At yet larger couplings, $Y(T)$ follows its equilibrium value $Y_{\rm eq}(T)$ for an extended period of time, before freeze-out. This signals DM thermalization.

We note that, in order to compute  $\Gamma ( \bar \chi \chi \rightarrow  {\rm SM} ) $, one needs to know the momentum distribution of dark matter. This reaction becomes important at large enough coupling and, thus, it is safe to assume that the SM-DM system is in kinetic equilibrium at that stage. 
   Indeed, the  elastic scattering 
reaction        SM+$\chi \rightarrow $ SM+$\chi$        is more efficient than the annihilation one due to the higher density of the SM states. Hence, kinetic equilibrium sets in at smaller couplings.  This is also the assumption adopted in {\tt micrOMEGAs} .

\section{Constraints and parameter space analysis}

In this section, we delineate parameter space of the model  producing the correct DM relic abundance and discuss various constraints on the ``universal''  coupling $\lambda$. These include  bounds on the $Z^\prime$ interactions  from collider experiments as well as constraints  on dark matter from direct and indirect searches.

\begin{figure}[h!]
    \centering
    \includegraphics[width=0.49\textwidth]{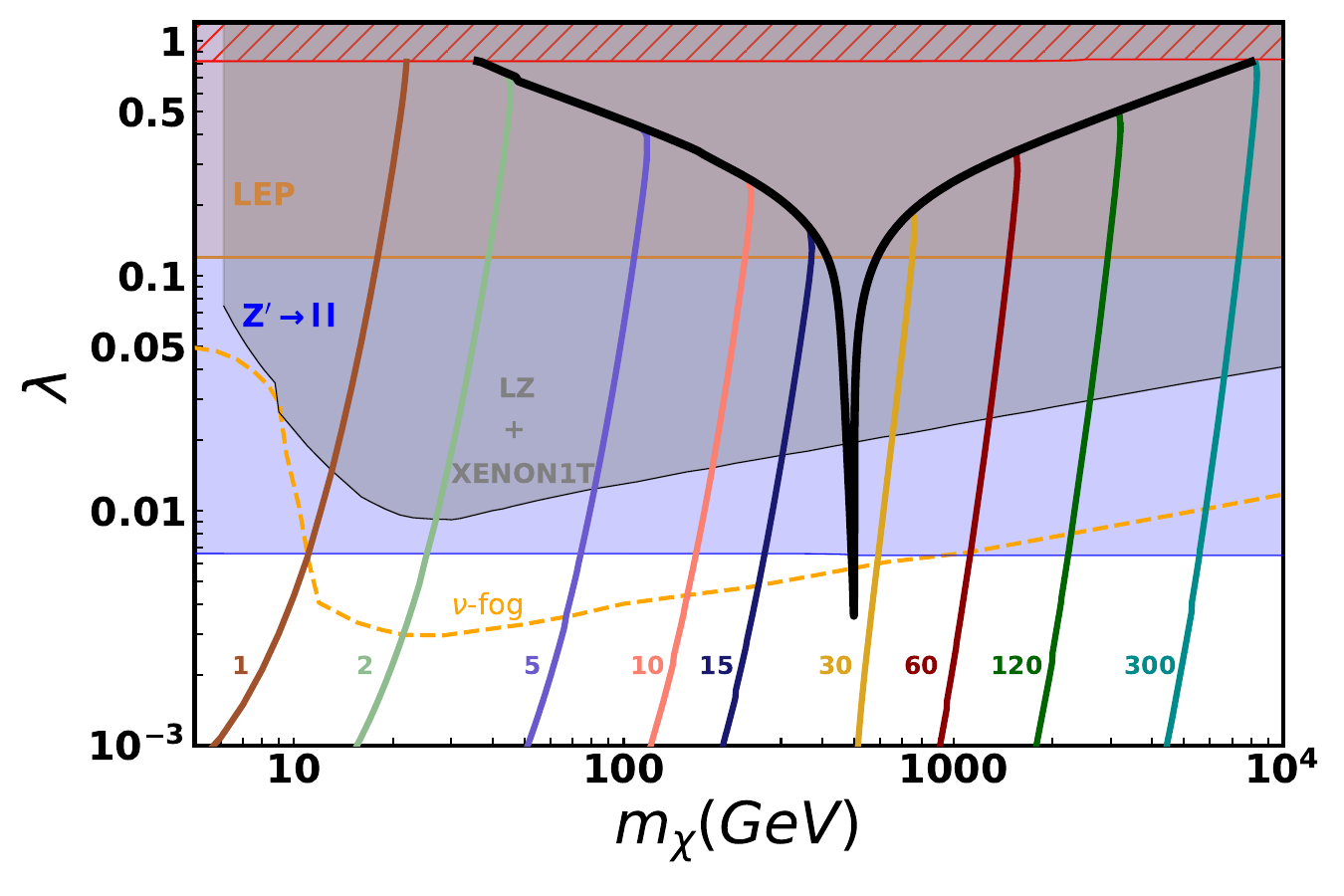}
    \includegraphics[width=0.49\textwidth]{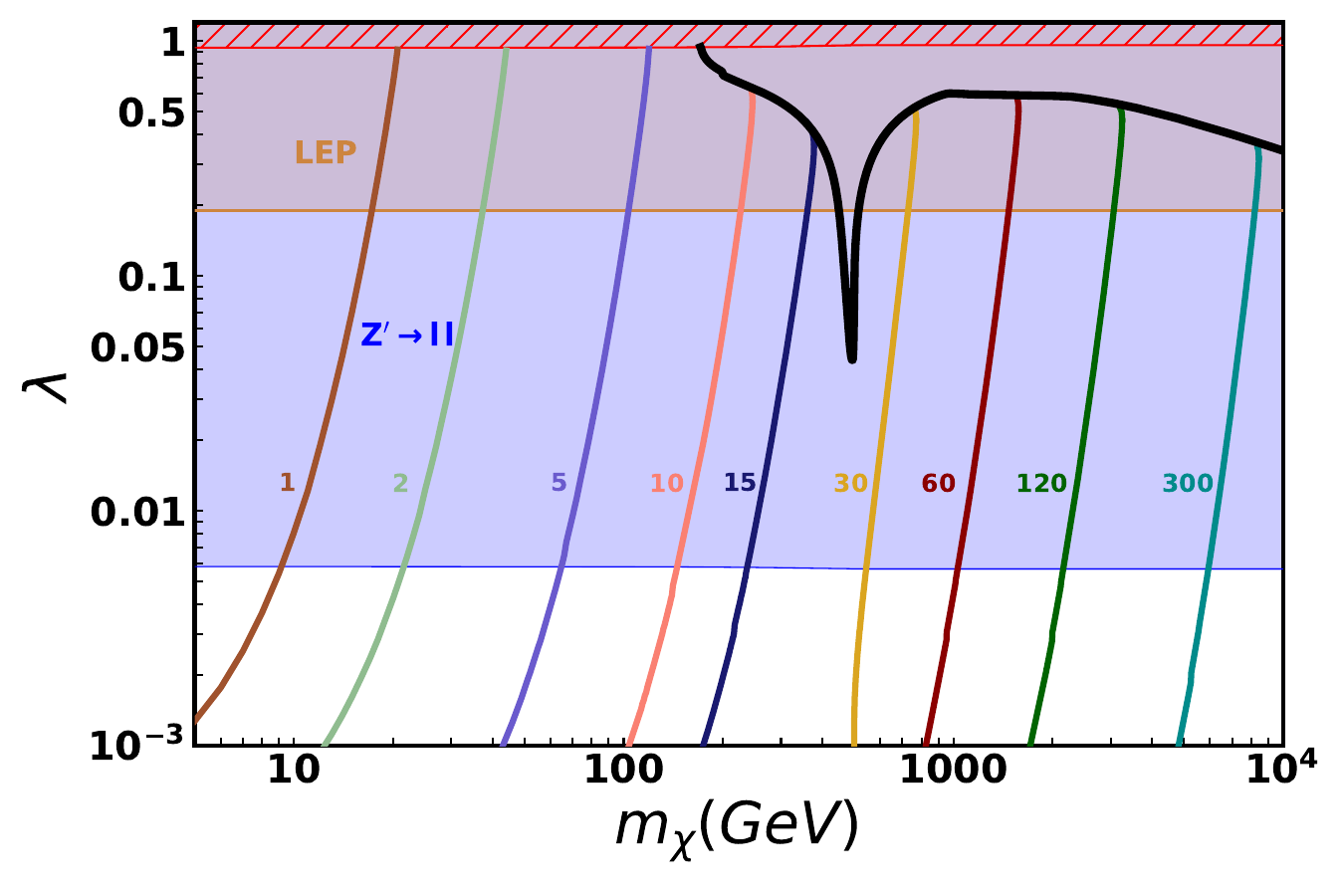}
    \includegraphics[width=0.49\textwidth]{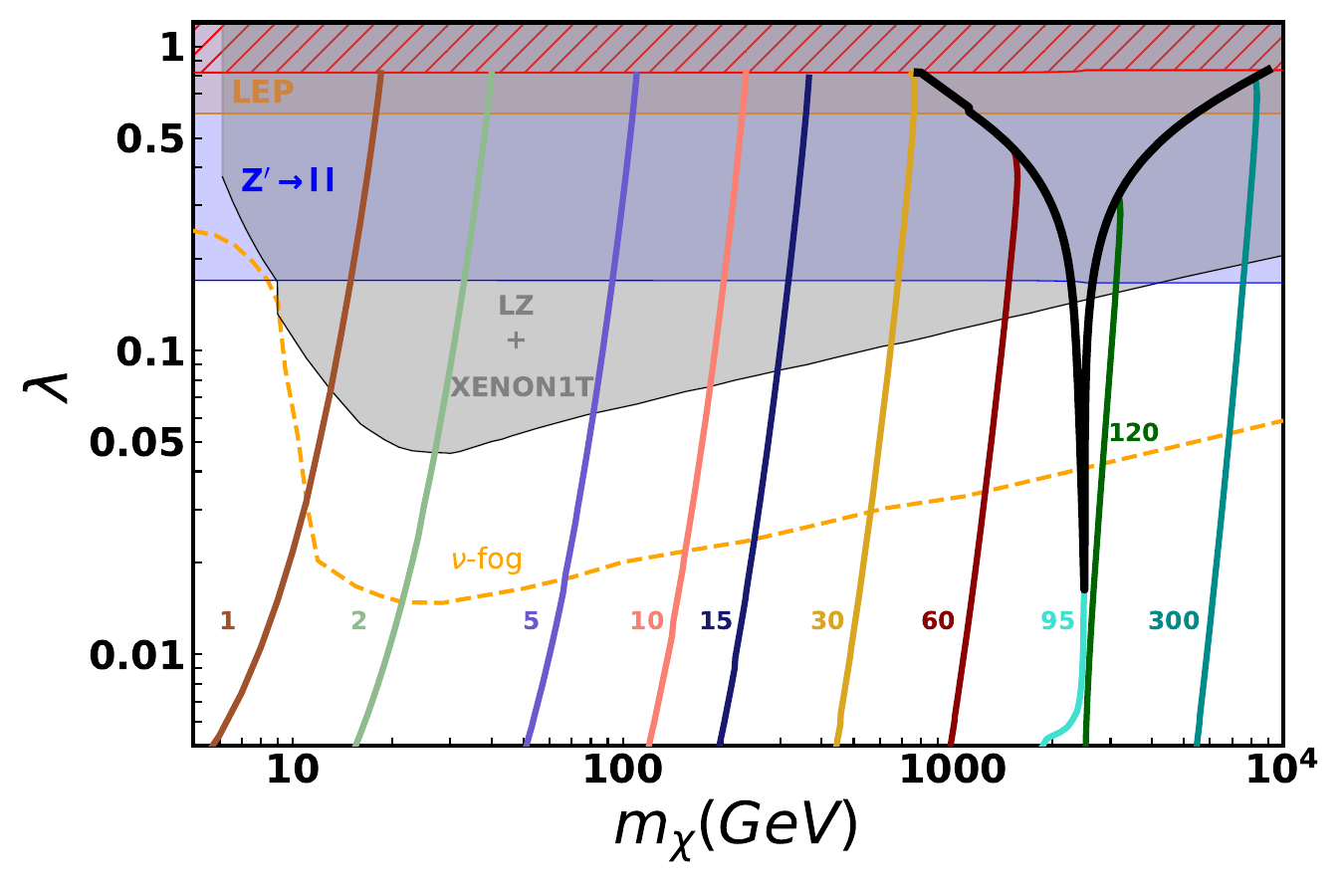}
    \includegraphics[width=0.49\textwidth]{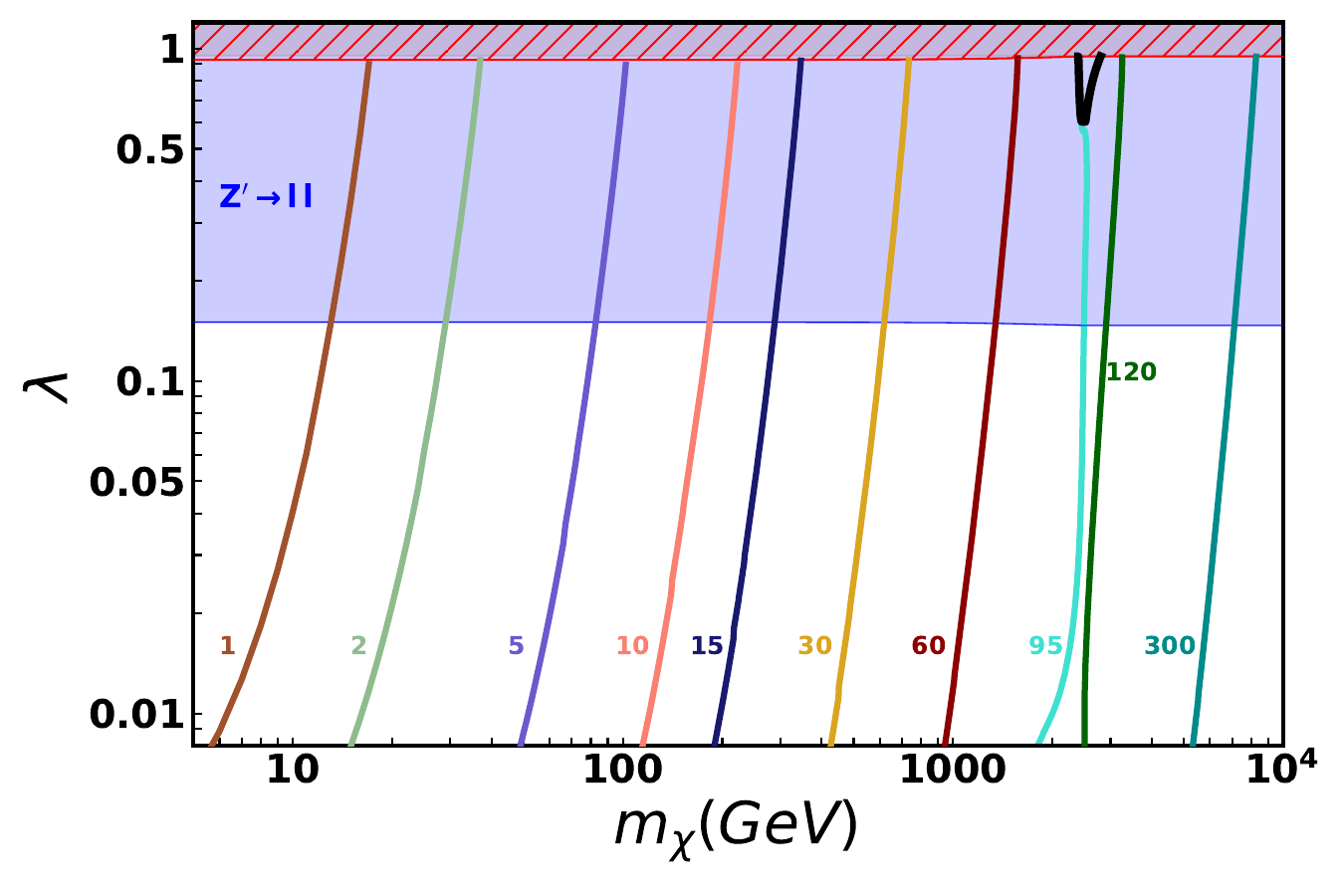}
     \includegraphics[width=0.49\textwidth]{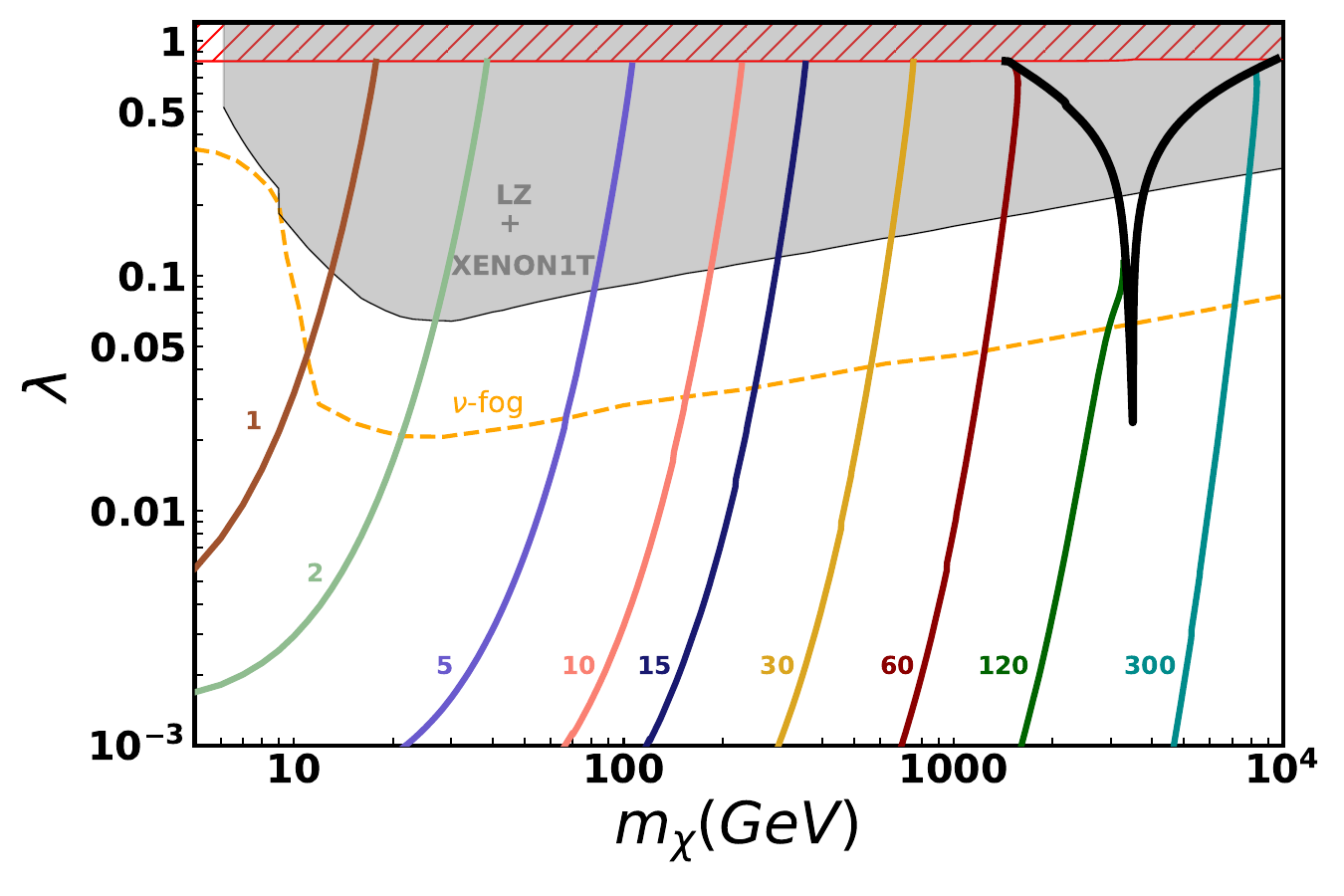}
    \includegraphics[width=0.49\textwidth]{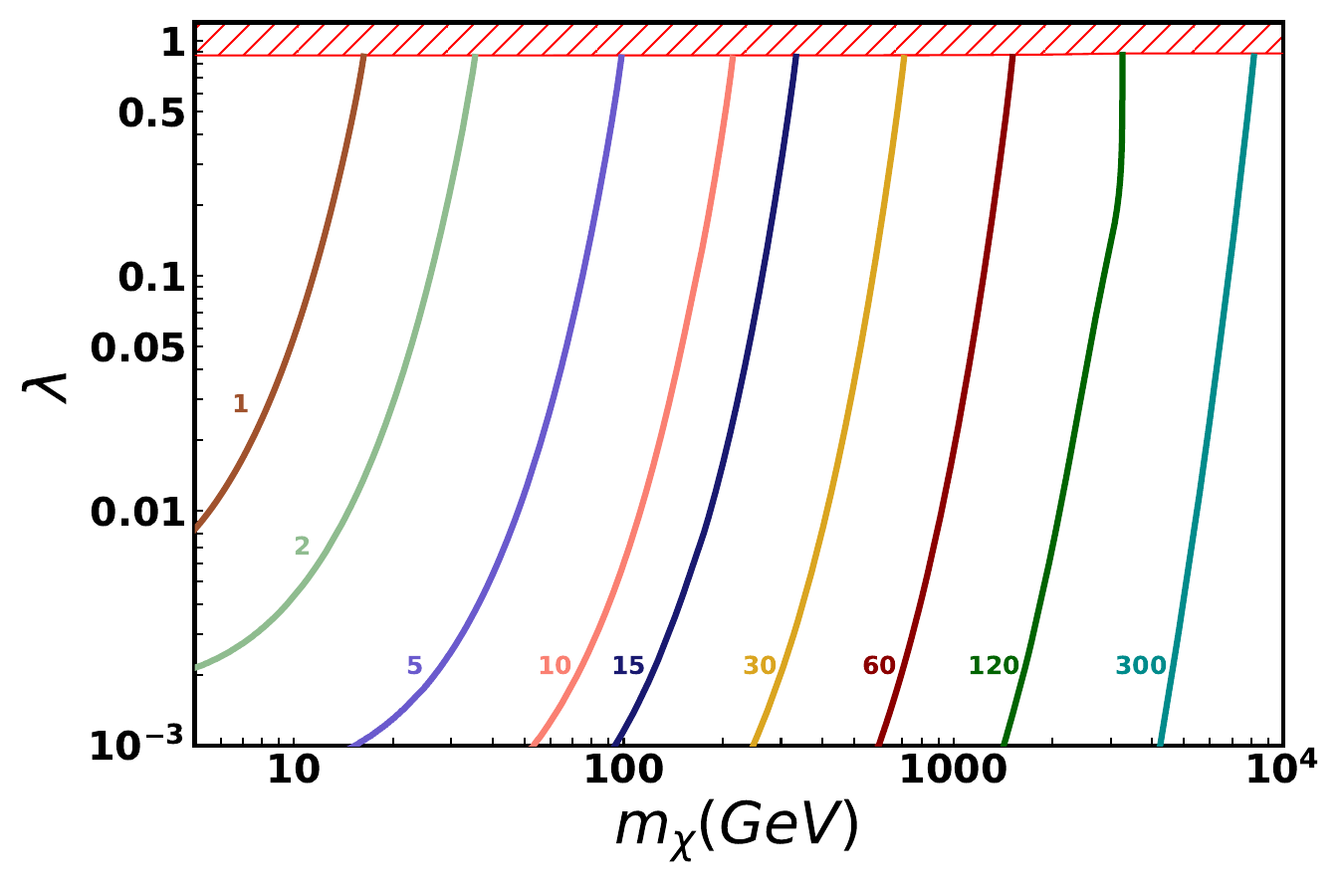}
    \caption{Allowed   vector and axial $Z^\prime$ couplings for different $M_{Z'}$. {\it Top}:  $M_{Z'}=1\;$TeV;   {\it Middle}:  $M_{Z'}=5\;$TeV; {\it  Bottom}:  $M_{Z'}=7\;$TeV. The $left$ panels correspond to the vector couplings 
    $V_f = V_\chi =\lambda$ (and $A_f = A_\chi =0$) , while the $right$ panels correspond to the axial couplings $A_f = A_\chi =\lambda$ (and $V_f = V_\chi =0$).
    Along the colored curves labelled by $T_R$ (in GeV), the correct DM relic density is reproduced. The black line corresponds to thermal DM undergoing freeze-out.
    The shaded areas are excluded and the ``$\nu$-fog'' curve represents the neutrino background for the direct DM detection experiments. }
    \label{par-space}
\end{figure}

\subsection{LEP constraints}

  A heavy $Z^\prime$ can be integrated out leading to a set of 4-fermion interactions. 
 The strictest constraints on the contact interactions are   imposed by the 4-lepton operators $\ell \ell \ell \ell$, while the lepton-quark operators    $\ell \ell  q q$  \cite{Cheung:2001wx} are less important.
 The most significant observables are the muon decay constant, which is affected by the muon-electron operators, as well as the LEP lepton measurements \cite{ALEPH:2006bhb,ALEPH:2013dgf}. 
 The relevant constraints for the flavor-universal case can be read off from Fig.\,3 (right) of \cite{Falkowski:2015krw}. 
 The vector coupling     ($\kappa_L=\kappa_R$)    is constrained by
 \begin{equation}
 {\lambda\over M_{Z'}} < {0.12 \over {\rm TeV}}
 \end{equation} 
 at 95 \% CL. The constraint on the axial coupling ($\kappa_L=-\kappa_R$) is somewhat weaker,
  \begin{equation}
 {\lambda\over M_{Z'}} < {0.19 \over {\rm TeV}}
 \end{equation} 
 at 95 \% CL.
The corresponding excluded regions  are shown in Fig.\,\ref{par-space}  in light orange, marked ``LEP''.
   
   \subsection{LHC constraints}
   
   The LHC experiments are conducting direct searches for a new gauge boson $Z^\prime$ in the channel $pp \rightarrow \ell^+ \ell^- X$, 
    where $X$ represents
the beam fragment jets. The constraints are obtained on the quantity 
    $\sigma (pp \rightarrow Z^\prime) \, {\rm BR}(Z^\prime \rightarrow \ell^+ \ell^- )$, employing
      the narrow width approximation. We use the current data from ATLAS \cite{ATLAS:2019erb} based on LHC Run 2 at $\sqrt{s}=13\;$TeV  
   with a total integrated luminosity of $L = 139\;{\rm fb}^{-1}$.
 The theory prediction for 
   $\sigma (pp \rightarrow Z^\prime)$ is  computed using 
  MadGraph \cite{Alwall:2014hca}. As the first step, we reproduce the ATLAS limits on   the $Z_\psi^\prime$ model \cite{ATLAS:2019erb}. Then, we derive the bounds on the universal coupling
   $\lambda$ in our model. The results are presented in  Fig.\,\ref{par-space}: the light blue area marked $Z^\prime \rightarrow \ell \ell$ is excluded by the LHC.
   
   The typical bound on the $Z^\prime$ mass for the SM-like couplings is around 5 TeV \cite{ATLAS:2019erb,Bandyopadhyay:2018cwu}. The constraint fades away very fast as $M_{Z'}$ increases and above 6-7 TeV the LHC 
    has no sensitivity to the extra gauge boson, at least at the perturbative level. This is unlike the other bounds (LEP, direct DM detection) which scale as a function of  $\lambda/M_{Z'}$.

    \subsection{Direct DM detection}

    Direct DM detection experiments traditionally set the strongest constraints on dark matter models (see, e.g. a recent discussion in \cite{Arcadi:2024ukq}). These are based on non-observation of any significant interaction of dark matter with nucleons. 
    In our case, such an interaction is mediated by the $Z^\prime$.
    The DM-nucleon scattering occurs at low energy, hence the $Z^\prime$ can be integrated out resulting in the effective Lagrangian 
  \begin{eqnarray}
  \mathcal{L}_{\text {eff }} & \supset &\frac{1}{M_{Z^{\prime}}^2}\left[V^\chi\left(2 V^u+V^d\right)\, \bar{\chi} \gamma^\mu \chi \;\bar{p} \gamma_\mu p\right. \\
&& \left.+\; A^\chi\left(\Delta_u^p \,A^u+\left(\Delta_d^p+\Delta_s^p\right) A^d\right) \bar{\chi} \gamma^\mu \gamma^5 \chi \; \bar{p} \gamma_\mu \gamma^5 p\right],
  \end{eqnarray}
   where $p$ represents the proton, and a similar expression applies to the neutron $n$. $\Delta_i^{p,n}$ stands for the spin content of  quark $i$ in the proton/neutron.
   According to \cite{HERMES:2006jyl},
\begin{equation}
\Delta_u^p=0.842, \quad \Delta_d^p=-0.427, \quad \Delta_s^p=-0.085 \;,
\end{equation}
and the corresponding neutron quantities are obtained by the isospin symmetry. 
 The resulting cross section for spin-independent scattering of DM on a nucleus with charge $Z$ and atomic weight $A$ is
 \begin{equation}
\sigma_{\chi N}^{\mathrm{SI}}=\frac{  \mu_{\chi N}^2 \left(  \,V^\chi\right)^2 }{\pi M_{Z^{\prime}}^4}\left[V^u\left(1+\frac{Z}{A}\right)+V^d\left(2-\frac{Z}{A}\right)\right]^2 \;,
\end{equation}  
  where $\mu_{\chi N}$  is the reduced mass of the DM-nucleon system. 
  This cross section must lie below the LZ 2022  and XENON1T bounds \cite{LZ:2022lsv,XENON:2018voc}, which imposes a significant constraint on the model parameters.
 For the case of the universal coupling $\lambda$, the direct detection constraint scales as 
  $\lambda^4/M_{Z'}^4$, while $\sigma_{\chi N}^{\mathrm{SI}}$ remains independent of the DM mass as long as $m_\chi \gg 1\;$GeV.
  
  The axial couplings lead to the spin-dependent DM-nucleon scattering, which is only weakly constrained
  and the resulting bounds are unimportant for our analysis.  The LHC searches impose the leading constraints in this case, at least for the $M_{Z'}$ range of interest.

  In Fig.\,\ref{par-space}, we show the  direct detection constraints on the universal coupling $\lambda$ with the help of  the {\tt micrOMEGAs} tool.  
  The grey regions marked by ``LZ+XENON1T'' are excluded.\footnote{ The  LZ collaboration has updated its 2022 direct detection bound. According to the preliminary (unpublished) estimate, the bound has improved by a factor of 5 - 10, depending on the DM mass. This translates into a stronger constraint on $\lambda$ by a factor of 1.5 - 1.8.}
  We observe that for a heavy $Z^\prime$, these constraints supersede the other bounds, while for a lighter  $Z^\prime$, the LHC
searches  set the strictest bounds. 
The figures also display the ``neutrino fog''  which represents the neutrino background for the direct DM detection experiments \cite{Billard:2021uyg}. The standard techniques would not be able to distinguish the DM scattering from that of neutrinos,
if $\lambda$ falls below the neutrino fog curve.  While the corresponding parameter space could be explored with innovative techniques, DM detection is challenging for such low couplings. 
Nevertheless, we see that large regions of the freeze-in parameter space, especially for a heavy $Z^\prime$, lie above the neutrino background and thus can be probed by the current and future direct DM detection experiments
such as XENONnT \cite{XENON:2020kmp} and DARWIN \cite{DARWIN:2016hyl}.\footnote{This property is shared by a class of  freeze-in models  with a light mediator \cite{Hambye:2018dpi,Boddy:2024vgt}
as well as the Higgs-portal-type  WIMP models with a first order phase transition \cite{Wong:2023qon}.  }

 \subsection{Indirect DM detection}
 
 Indirect dark matter detection is based on possible signatures of its annihilation in regions with significant DM density, e.g. the Galactic center. 
 The strictest bounds come from the 
  Fermi-LAT   \cite{McDaniel:2023bju} observation of 30 dSphs for 14.3 years.
  Interpreting these constraints in our model using  {\tt micrOMEGAs}, we find that they are superseded by the bounds discussed above and thus are insignificant.
   
  \subsection{Discussion and summary}
  
 The correct DM relic abundance is reproduced 
  along the colored lines in Fig.\,\ref{par-space}. Each of them has a different reheating temperature $T_R$, which we also identify with the maximal SM bath temperature.  
  We observe that, at sufficiently large coupling, all  the freeze-in    lines   merge with the thermal  black curve. This signals DM thermalization such that the relic abundance is controlled by the 
  usual freeze-out. The corresponding parameter space is, however, ruled out experimentally, leaving only a very narrow resonance region $m_\chi \simeq M_{Z'}/2$.

  In addition to the  constraints discussed above, we impose the perturbativity bound which excludes the red hatched area.
    In this   region, the $Z^\prime$ resonance becomes too broad such that $\Gamma_{Z'} > 0.5 \; M_{Z'}$ and it ceases to be a ``particle'' in the conventional sense.
  While for a light $Z^\prime$ this bound is superseded by other constraints, in the case of a heavy $Z^\prime$, it becomes important, especially for the axial-type couplings.

   The axial coupling case is constrained mostly by collider observables, whereas  the direct detection bounds are very weak. This also implies that an axial $Z^\prime$-induced freeze-in can hardly be probed experimentally.
   On the other hand, the vector case is more interesting and significant parts of the freeze-in parameter space can be probed by the direct detection experiments. 
   For $M_{Z'}=1\;$TeV, the LHC constraints dominate, yet they leave a modestly sized region above the $\nu$-fog line, which can therefore be probed by direct DM detection. The size of this area grows as $M_{Z'}$ increases
   and at $M_{Z'}=7\;$TeV, the LHC bounds fade away. 
     
    We observe that, as long as the $Z^\prime$ couplings have a significant vector component, the direct DM detection prospects are good in the entire range of the DM mass considered, $10-10^4\,$ GeV.  
     The required gauge coupling is in the range      $10^{-3}-10^{-1}$ for a $Z^\prime$ mass between 1 and 10 TeV.

\section{Conclusions}
   
   We have analyzed $Z^\prime$-mediated fermion dark matter production in the freeze-in regime at stronger coupling, when the SM bath temperature is below the DM mass.
   Models with a low reheating temperature are motivated by the problem of gravitational particle production which mars  the usual freeze-in mechanism. In particular, Planck--suppressed
   operators coupling dark matter to the inflaton are very efficient in particle production immediately after inflation, which leads to a large initial abundance of dark matter. This problem can be solved by 
   allowing for an extended  period of inflaton-dominated expansion thereby diluting the initial DM abundance.  As a result, the reheating temperature in this framework  is relatively low, depending on further details.
 
   Using both analytic estimates and more sophisticated numerical tools, we find that the correct DM relic abundance can be produced for a broad range of the DM mass $m_\chi$, assuming a TeV-scale ${Z^\prime}$. The main factor is the reheating temperature
   $T_R$ and, for $m_\chi / T_R \sim 20$, the $Z^\prime $ couplings can be as large as ${\cal O}(1)$.  We distinguish the vector and axial coupling cases, which exhibit different phenomenology. While the DM production rates are similar in both cases, 
   the   constraints on the parameter space differ substantially.   In particular, the direct DM detection results  set strict bounds on  the vector coupling, whereas the axial coupling remains essentially unconstrained. 
   In our freeze-in framework, this implies that the vector case can be probed by the current and future direct detection experiments, which will reach the sensitivity at the level of the ``neutrino fog''.  The axial case, on the other hand, is very difficult to test due to the suppression of the DM-nucleus cross section. Further bounds on the parameter space are imposed by the direct $Z^\prime$ searches at the LHC, LEP measurements of the lepton production and
   constraints from the muon decay. 
   
   As the $Z^\prime$ coupling increases, the DM production grows more efficient and the inverse reaction becomes significant.  At some critical value of the coupling (depending on $T_R$), both reactions equilibrate and dark matter thermalizes. 
   We observe this explicitly as the freeze-in relic abundance curves merge with the standard freeze-out line. In the vector coupling case, the corresponding region of the parameter space is ruled out by the direct DM detection.
   However, at lower couplings, the Boltzmann-suppressed freeze-in is still operative and produces the correct relic abundance while evading 
    such constraints. For a range of the vector couplings, typically of order $10^{-3}-10^{-1}$, the DM-nucleon cross section lies above the neutrino fog and thus can be tested in the near future.  Needless to say, this does not require the $Z^\prime$
to be purely ``vectorial'', it only sets a lower bound on the vector component of the $Z^\prime$ coupling. 
    
    Our main conclusion is that, as long as the vector coupling of the $Z^\prime$ is substantial,  the  freeze-in dark matter can be probed further (and possibly discovered) by direct DM detection experiments such as XENONnT and DARWIN.  
   \\ \ \\
   {\bf Acknowledgments}
   \newline
   The authors thank Francesco Costa for fruitful discussions.
   D.C.A. acknowledges funding from the Spanish MCIN/AEI/10.13039/501100011033 through grant PID2022-136224NB-C21.

   \appendix
\section{General cross sections and decay widths }   
   
   In this appendix, we provide some analytical formulas for the general case of the vector and axial couplings being present simultaneously. 
   The DM annihilation cross section 
    into a Standard Model Dirac fermion $f$ is given by   
\begin{eqnarray}
\sigma_{\bar \chi \chi \to \bar f f}&=&
\nonumber
\frac{1}{12 \pi s\left[\left(s-M_{Z^\prime}^2\right)^2+M_{Z^\prime}^2 \Gamma_{Z'}^2\right]} \sqrt{\frac{1-4 m_f^2 / s}{1-4 m_\chi^2 / s}}\left[A_ {f} ^ { 2 }V_\chi^2\left(s-4 m_f^2\right)\left(2 m_\chi^2+s\right)\right. \\
&&+A_ {f}^{2}A_\chi^2\left(4 m_\chi^2\left[m_f^2\left(7-\frac{6 s}{M_{Z^\prime}^2}+\frac{3 s^2}{M_{Z^\prime}^4}\right)-s\right]
+s\left(s-4 m_f^2\right)\right)\nonumber\\
 &&\left.+V_f^2\left(2 m_f^2+s\right)\left(A_\chi^2\left(s-4 m_\chi^2\right)+V_\chi^2\left(2 m_\chi^2+s\right)\right)\right]\;,
 \end{eqnarray}
 where $s$ is the Mandelstam variable. This result agrees with the corresponding cross section presented in \cite{Berlin:2014tja}.
  The $Z^\prime$ decay width is
\begin{equation}
\Gamma_{Z'}=\Gamma_{Z^{\prime} \to \bar\chi \chi  }+\sum_{f}\Gamma_{Z^{\prime} \to \bar f f} \; ,
\label{eq:gammav}
\end{equation}
with
\begin{equation}
\scalebox{1.2}{$\Gamma_{Z^{\prime} \to \bar \chi\chi}=\frac{\sqrt{M_{Z'}^2-4 m_{\chi }^2} \;\left(A_{\chi }^2 \left(M_{Z'}^2-4 m_{\chi }^2\right)+V_{\chi
   }^2 \left(2 m_{\chi }^2+M_{Z'}^2\right)\right)}{12 \pi  M_{Z'}^2} \;,$}  
\label{eq:gammaX}
 \end{equation}
\begin{equation}
\scalebox{1.2}{$\Gamma_{Z^{\prime}\to \bar f f}=\frac{ \sqrt{M_{Z'}^2-4 m_f^2} \; \left(A_f^2 \left(M_{Z'}^2-4 m_f^2\right)+V_f^2 \left(2
   m_f^2+M_{Z'}^2\right)\right)}{12 \pi  M_{Z'}^2} \;.$}
\label{eq:gammaf}
 \end{equation}
 In the universal coupling case with a heavy $Z^\prime$, the sum over the SM fermions amounts to a factor of 22.5, which includes 3$\times$6 quark contributions, $3$ charged lepton contributions and 3$\times$1/2 neutrino terms.  
   The branching ratio for the $Z^{\prime} \to e^+ e^-$ decay approaches 1/23.5 in this limit.

The $t$-channel $Z^\prime Z^\prime \rightarrow \bar \chi \chi $ analytical results can be found in \cite{Klasen:2016qux}, which we have also reproduced.


\begin{thebibliography}{99}


\bibitem{Dodelson:1993je} 
  S.~Dodelson and L.~M.~Widrow,
  Phys.\ Rev.\ Lett.\  {\bf 72}, 17 (1994).
  


\bibitem{Hall:2009bx}
L.~J.~Hall, K.~Jedamzik, J.~March-Russell and S.~M.~West,
JHEP \textbf{03}, 080 (2010).




\bibitem{Lebedev:2022cic}
O.~Lebedev,
JCAP \textbf{02}, 032 (2023).



\bibitem{Lebedev:2022ljz}
O.~Lebedev and J.~H.~Yoon,
JCAP \textbf{07}, no.07, 001 (2022).


\bibitem{Koutroulis:2023fgp}
F.~Koutroulis, O.~Lebedev and S.~Pokorski,
JHEP \textbf{04}, 027 (2024).



\bibitem{Hannestad:2004px}
S.~Hannestad,
Phys. Rev. D \textbf{70} (2004), 043506.



\bibitem{Cosme:2023xpa}
C.~Cosme, F.~Costa and O.~Lebedev,
Phys. Rev. D \textbf{109}, no.7, 075038 (2024).



\bibitem{Langacker:2008yv}
P.~Langacker,
Rev. Mod. Phys. \textbf{81} (2009), 1199-1228






\bibitem{Arcadi:2013qia}
G.~Arcadi, Y.~Mambrini, M.~H.~G.~Tytgat and B.~Zaldivar,
JHEP \textbf{03} (2014), 134.

\bibitem{Lebedev:2014bba}
O.~Lebedev and Y.~Mambrini,
Phys. Lett. B \textbf{734} (2014), 350-353.

\bibitem{Arcadi:2014lta}
G.~Arcadi, Y.~Mambrini and F.~Richard,
JCAP \textbf{03} (2015), 018




\bibitem{Cosme:2024ndc}
C.~Cosme, F.~Costa and O.~Lebedev,
JCAP \textbf{06} (2024), 031.





\bibitem{Cosme:2021baj}
C.~Cosme, M.~Dutra, S.~Godfrey and T.~R.~Gray,
JHEP \textbf{09} (2021), 056



















 
\bibitem{Gondolo:1990dk}
P.~Gondolo and G.~Gelmini,
Nucl. Phys. B \textbf{360} (1991), 145-179



\bibitem{Koivunen:2024vhr}
N.~Koivunen, O.~Lebedev and M.~Raidal,
[arXiv:2403.15533 [hep-ph]].

\bibitem{Arcadi:2024wwg}
G.~Arcadi, F.~Costa, A.~Goudelis and O.~Lebedev,
JHEP \textbf{07} (2024), 044.




\bibitem{Belanger:2018ccd}
G.~B\'elanger, F.~Boudjema, A.~Goudelis, A.~Pukhov and B.~Zaldivar,
Comput. Phys. Commun. \textbf{231}, 173-186 (2018).

\bibitem{Alguero:2023zol}
G.~Alguero, G.~Belanger, F.~Boudjema, S.~Chakraborti, A.~Goudelis, S.~Kraml, A.~Mjallal and A.~Pukhov,
Comput. Phys. Commun. \textbf{299}, 109133 (2024).





\bibitem{Silva-Malpartida:2024emu}
J.~Silva-Malpartida, N.~Bernal, J.~Jones-P\'erez and R.~A.~Lineros,
[arXiv:2408.08950 [hep-ph]].












\bibitem{Cheung:2001wx}
K.~m.~Cheung,
Phys. Lett. B \textbf{517} (2001), 167-176

 
 
\bibitem{ALEPH:2006bhb}
J.~Alcaraz \textit{et al.} [ALEPH, DELPHI, L3, OPAL and LEP Electroweak Working Group],
[arXiv:hep-ex/0612034 [hep-ex]].

\bibitem{ALEPH:2013dgf}
S.~Schael \textit{et al.} [ALEPH, DELPHI, L3, OPAL and LEP Electroweak],
Phys. Rept. \textbf{532} (2013), 119-244



\bibitem{Falkowski:2015krw}
A.~Falkowski and K.~Mimouni,
JHEP \textbf{02} (2016), 086





\bibitem{ATLAS:2019erb}
G.~Aad \textit{et al.} [ATLAS],
Phys. Lett. B \textbf{796} (2019), 68-87


\bibitem{Alwall:2014hca}
J.~Alwall, R.~Frederix, S.~Frixione, V.~Hirschi, F.~Maltoni, O.~Mattelaer, H.~S.~Shao, T.~Stelzer, P.~Torrielli and M.~Zaro,
JHEP \textbf{07} (2014), 079.


\bibitem{Bandyopadhyay:2018cwu}
T.~Bandyopadhyay, G.~Bhattacharyya, D.~Das and A.~Raychaudhuri,
Phys. Rev. D \textbf{98} (2018) no.3, 035027





\bibitem{Arcadi:2024ukq}
G.~Arcadi, D.~Cabo-Almeida, M.~Dutra, P.~Ghosh, M.~Lindner, Y.~Mambrini, J.~P.~Neto, M.~Pierre, S.~Profumo and F.~S.~Queiroz,
[arXiv:2403.15860 [hep-ph]].



\bibitem{HERMES:2006jyl}
A.~Airapetian \textit{et al.} [HERMES],
Phys. Rev. D \textbf{75} (2007), 012007

 
 
 
 
\bibitem{LZ:2022lsv}
J.~Aalbers \textit{et al.} [LZ],
Phys. Rev. Lett. \textbf{131} (2023) no.4, 041002


\bibitem{XENON:2018voc}
E.~Aprile \textit{et al.} [XENON],
Phys. Rev. Lett. \textbf{121}, no.11, 111302 (2018).




\bibitem{Billard:2021uyg}
J.~Billard, M.~Boulay, S.~Cebri\'an, L.~Covi, G.~Fiorillo, A.~Green, J.~Kopp, B.~Majorovits, K.~Palladino and F.~Petricca, \textit{et al.}
Rept. Prog. Phys. \textbf{85}, no.5, 056201 (2022).



\bibitem{XENON:2020kmp}
E.~Aprile \textit{et al.} [XENON],
JCAP \textbf{11}, 031 (2020).
 
   
\bibitem{DARWIN:2016hyl}
J.~Aalbers \textit{et al.} [DARWIN],
JCAP \textbf{11}, 017 (2016).


 \bibitem{Hambye:2018dpi}
T.~Hambye, M.~H.~G.~Tytgat, J.~Vandecasteele and L.~Vanderheyden,
Phys. Rev. D \textbf{98}, no.7, 075017 (2018).


\bibitem{Boddy:2024vgt}
K.~K.~Boddy, K.~Freese, G.~Montefalcone and B.~Shams Es Haghi,
[arXiv:2405.06226 [hep-ph]].


\bibitem{Wong:2023qon}
X.~R.~Wong and K.~P.~Xie,
Phys. Rev. D \textbf{108} (2023) no.5, 055035








 

\bibitem{McDaniel:2023bju}
A.~McDaniel, M.~Ajello, C.~M.~Karwin, M.~Di Mauro, A.~Drlica-Wagner and M.~A.~S\'anchez-Conde,
Phys. Rev. D \textbf{109} (2024) no.6, 063024




 
 
 

\bibitem{Berlin:2014tja}
A.~Berlin, D.~Hooper and S.~D.~McDermott,
Phys. Rev. D \textbf{89} (2014) no.11, 115022

 
\bibitem{Klasen:2016qux}
M.~Klasen, F.~Lyonnet and F.~S.~Queiroz,
Eur. Phys. J. C \textbf{77} (2017) no.5, 348

 



\end{thebibliography}
\end{document}